\def \SAIT #1 #2 {{\em Mem.\ Soc.\ Astron.\ It.\/} {\bf #1}, #2}
\def \MESS #1 #2 {{\em The Messenger\/} {\bf #1}, #2}
\def \ASTRNACH #1 #2 {{\em Astron. Nach.\/} {\bf #1}, #2}
\def \AAP #1 #2 {{\em Astron. Astrophys.\/} {\bf #1}, #2}
\def \AAL #1 #2 {{\em Astron. Astrophys. Lett.\/} {\bf #1}, L#2}
\def \AAR #1 #2 {{\em Astron. Astrophys. Rev.\/} {\bf #1}, #2}
\def \AAS #1 #2 {{\em Astron. Astrophys. Suppl. Ser.\/} {\bf #1}, #2}
\def \AJ #1 #2 {{\em Astron. J.\/} {\bf #1}, #2}
\def \ANNREV #1 #2 {{\em Ann. Rev. Astron. Astrophys.\/} {\bf #1}, #2}
\def \APJ #1 #2 {{\em Astrophys. J.\/} {\bf #1}, #2}
\def \APJL #1 #2 {{\em Astrophys.. J. Lett.\/} {\bf #1}, L#2}
\def \APJS #1 #2 {{\em Astrophys. J. Suppl.\/} {\bf #1}, #2}
\def \APSS #1 #2 {{\em Astrophys. Space Sci.\/} {\bf #1}, #2}
\def \ASR #1 #2 {{\em Adv. Space Res.\/} {\bf #1}, #2}
\def \BAIC #1 #2 {{\em Bull. Astron. Inst. Czechosl.\/} {\bf #1}, #2}
\def \JSQRT #1 #2 {{\em J. Quant. Spectrosc. Radiat. Transfer\/} {\bf #1}, #2}
\def \MN #1 #2 {{\em Mon. Not. R. Astr. Soc.\/} {\bf #1}, #2}
\def \MEM #1 #2 {{\em Mem. R. Astr. Soc.\/} {\bf #1}, #2}
\def \PLR #1 #2 {{\em Phys. Lett. Rev.\/} {\bf #1}, #2}
\def \PASJ #1 #2 {{\em Publ. Astron. Soc. Japan\/} {\bf #1}, #2}
\def \PASP #1 #2 {{\em Publ. Astr. Soc. Pacific\/} {\bf #1}, #2}
\def \NAT #1 #2 {{\em Nature\/} {\bf #1}, #2}

\documentstyle{memsait}
\input epsf.sty
\begin{opening}
\title{IF THE MACHOS ARE OLD WHITE DWARFS........}
\author{Harvey B. Richer$^{1,2}$}
\institute{$^1$Osservatorio Astronomico di Roma, Monteporzio Catone, Roma, Italia}
\institute{$^2$Department of Physics and Astronomy, University of British Columbia, Vancouver, Canada\\
}
\date{} 
\end{opening}

\begin{document}

\oddpagefooter{}{}{} 
\evenpagefooter{}{}{} 
\ 
\bigskip

\begin{abstract}
The microlensing experiments in the direction of the LMC
seem to be indicating  
that about 60\% of the dark matter in the Galactic halo is tied
up in objects whose masses are about half the mass of the Sun. This mass is a natural one for
old white
dwarfs although other possibilities do exist. Using a grid of newly constructed
 models of
cooling white dwarfs which incorporate for the first time the effects of
molecular opacity in the stellar atmosphere, and assuming that such white
dwarfs make up the entire Galactic dark matter, I predict the numbers of old
white dwarfs expected in various surveys currently being conducted. In
particular, I note the number to be expected in the Hubble Deep Field (HDF), the
deepest image of the sky yet obtained.

\end{abstract}

\section{Background -- Can the MACHOs be Old White Dwarfs?}

The microlensing experiments in the direction of the LMC
seem to be indicating  
that $60 \pm 20$\% of the dark matter in the Galactic halo is tied
up in $0.5^{+0.3}_{-0.2} M_{\odot}$ objects (Alcock {\it et al.} 1997a, 1997b; Renault {\it et al.} 1997). This mass is a natural one for old white
dwarfs as this appears to be about the masses of white dwarfs found in globular
clusters (Richer {\it et al.} 1995, 1997) although other possibilities exist 
(e.g. neutron stars or primordial black holes). However, this scenario has
numerous problems associated with it. Some of these are listed below together
with their possible solutions.
\vspace{0.5cm}

\begin{enumerate}
\item{{\it Since these old white dwarfs represent roughly a quarter of the total
mass of the original star, where is the rest of this mass?} 

The
problem here is that if old white dwarfs represent about $2 \times 10^{11} M_{\odot}$
of material in the Galaxy, where is the other $ 6 \times 10^{11} M_{\odot}$ or so 
of gas ejected
when the precursors formed the white dwarfs? One answer may be that it was
 blown out of
the Galaxy early on by the first generation of supernovae. In a broader
context, we see x-ray gas in large clusters of galaxies and in filaments
between clusters and it seems
clear now that this represents a substantial amount of mass, 
rivalling that in the
cluster galaxies themselves. 
Further, much of this gas has been strongly enriched so it must have
been cycled through stars. This gas may represent the material ejected by the
stars from the initial burst of star formation in our Galaxy, the remnants 
from these stars, now in white dwarfs, making up the bulk of the Galactic dark
matter. Another possibility is that this helium-enriched gas could possibly
have condensed into dark clouds. Such clouds
would be very difficult to observe at any wavelength, even including CO.} 
\vspace{0.2in}
\item{{\it If there were so many white dwarfs formed as a result of 
the first burst
of star formation in our Galaxy, there must have been a huge tail of their
precursors that had low mass and these should be producing bright white dwarfs
in the halo today. Unfortunately, these bright white dwarfs are just not seen.} 

The
potential answer to this objection is contained in the work of Chabrier 1999, 
Gibson and Mould 1997, Chabrier {\it et al.} 1996, and Adams and Laughlin 1996. These authors have investigated
 IMFs with log-normal or truncated power-law
shapes  to avoid this problem and also solve the chemical evolution problem 
wherein the massive stars would overproduce heavy elements.
Such IMFs would have a peak near $2.5 M_{\odot}$, a virtual cut off at the high mass end near
$6 M_{\odot}$, and at low mass just below $2 M_{\odot}$. Almost no 
low mass stars (that would be making white dwarfs today) are thus present.}

\vspace{0.2in}

\item{{\it We don't see stellar systems anywhere in which stars are currently being formed with this peculiar IMF.} 

This appears to be true but neither
do we observe systems where the physical conditions are those of the early
Universe. Our ignorance of star formation theory is such that at this stage
such unusual IMFs can certainly not be excluded. It remains for theorists to try
and understand how such an IMF could have resulted from star formation shortly
after the Big Bang and for observers to try and find very metal-poor systems in
which the IMF can be determined.}

\vspace{0.2in}

\item{{ \it Calculations of the current chemical abundances
under the white dwarf hypothesis have indicated that the Galaxy should
be much more enriched in C and N than is
currently observed (Gibson and Mould 1997)}. 

There are two primary scenarios for avoiding C and N overproduction; either the gas is blown out by
supernovae or these heavy elements are just not produced in $z=0$ stars.  
The former hypothesis was the one favored by Fields {\it et
al.} 1997, but there are difficulties with the premise.
In their picture intermediate mass stars, regardless of $z$, undergo
thermal pulses and eject large amounts of C$^{12}$ and N$^{14}$, but Type II 
SNe going off
simultaneously blow everything out of the Galaxy into the inter-Local
Group medium.  One difficulty with this picture is that they had to adopt an
instantaneous recycling approximation (IRA) to ensure the SNe went off at the
same time as the AGB stars were returning ejecta, allowing a wind to develop
before the ejecta could get incorporated into any subsequent generation of
stars.  In reality though there is (roughly) an order of magnitude
difference in timescales for Type II SNe and thermally pulsing AGB stars, so the IRA is a poor
assumption to make here.

If WDs are the solution to the Galactic dark matter problem, then it is more likely that the second option will
prevail, 
i.e. primordial $z=0$ (population III) $1$ -- $8 M_{\odot}$  stars simply do not behave like
$[Fe/H] = -1.0$ stars. They may not undergo pulses and the third dredge-up and hence will
 not return much, if
anything, in the way of newly synthesized yields back to the
interstellar medium.  Even in the mid-1980s,
there was a hint of such a possibility in the work of 
Chieffi and Tornambe 1984 who showed that a $z=0$, $5 M_{\odot}$ star did not undergo thermal 
pulses. On the other hand, they inferred that all their $1$ -- $4 M_{\odot}$ models should go
through a thermally pulsing phase (but no detailed calculations were made),
 and for the IMFs we are considering here,
it is only the $1$ -- $4 M_{\odot}$ regime which is important. Still, this
 is perhaps a
hint that there is a way to produce huge numbers of population
 III intermediate mass stars without polluting the Galaxy too much. Clearly
much more theoretical work is required here.}

\vspace{0.2in}

\item{{\it Surely our Galaxy was not unique. Presumably, thus, most galaxies
in the early Universe would have undergone a phase of similar star formation. Are the
properties of high redshift galaxies consistent with this hypothesis?} 

Charlot and Silk 1995 examined this question and concluded from the number counts
of faint galaxies in deep surveys that current galactic
halos could not contain more than about $10\%$ of their mass in old white
 dwarfs formed from intermediate mass stars. It might be time to relook 
at these earlier calculations and include
the effects of strong dust absorption, different IMFs for the first 
generation of stars, and very low metallicity stellar evolutionary models
which could alter significantly the stellar lifetimes. New data in the submillimeter region of the
spectrum from SCUBA on the JCMT (Smail {\it et al.} 1999) is indicating that there may be an important
population of highly obscured young galaxies which could produce an important
upward revision to the $10\%$ figure.} 
\end{enumerate}

\section{New White Dwarf Cooling Models}

Any search for old white dwarfs must be guided by theory -- clearly we must
know what we are looking for and design the experiments accordingly. Until recently all cooling white dwarf models predicted that these old objects would
 continue to fade and get cooler and redder as time progressed. However, 
recently new models (Hansen
1998, 1999; Saumon and Jacobson 1999) were developed that include 
the effects of molecular opacity 
which is critical in understanding the luminosity and emergent spectrum of
white dwarfs whose temperatures fall below 4000K. Hydrogen
molecules provide
a very important opacity source which has the effect of redistributing
the emergent radiation from the infrared in to the blue so that {\it old
white dwarfs are blue and not red.}  

Why is this important? Two potential observables are affected by this
redistribution of energy; the colors and magnitudes of old white dwarfs.

 First, cool white dwarfs are blue and not red.
In deep images of the sky there are huge numbers of faint galaxies whose
predominant color is red. At faint limits it is usually impossible to distinguish
a star from an {\it almost} unresolved galaxy by image morphology alone. 
If old white dwarfs are blue they will move
away from the galaxies in an optical color-magnitude diagram and be much easier
to locate. For example, a typical faint galaxy has a $V-I$ color
of about $1.0$ whereas a 14 Gyr $0.5M_{\odot}$ white dwarf should have
$V - I = -0.8$. There are almost no other objects expected at this color.

Secondly, even though the bolometric magnitude of the white dwarf continues
to fall as it ages and cools, the absolute $V$ magnitude remains fairly constant at
about $M_V = 17.0$ as more and more of the stellar flux is squeezed in to
the $V$ band due to an $H_2$ opacity minimum in this wavelength range.

Both these effects can be seen in Figure 1 where I plot Hansen's cooling
model for a $0.5M_{\odot}$ white dwarf with a hydrogen-rich atmosphere
compared to an earlier model that does not include the molecular opacity.
The difference is quite dramatic and shows why
there might be some optimism that very old white dwarfs could be observed -- at least
compared to what one expected from the earlier models.

\begin{figure}[h]
\epsfxsize=7.5cm 
\hspace{3.5cm}\epsfbox{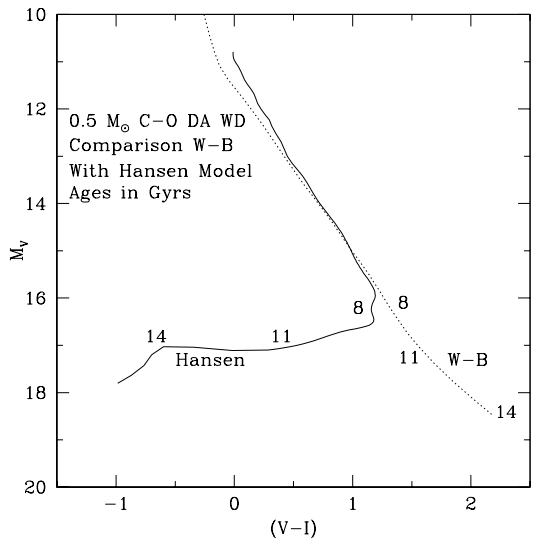} 
\caption[h]{A $0.5 M_{\odot}$ white dwarf cooling model of Hansen 1998, 1999 compared
with a similar mass model constructed from the interiors of Wood 1992, 1995 
and Bergeron {\it et al.} 1995 atmospheres (W-B). The age of the white dwarf 
is indicated on the diagram in Gyr. The main differences set in at around 
8 Gyr where the effects of atmospheric $H_2$ opacity become important.}
\end{figure}
\vspace{1cm}

\section{Number of White Dwarfs Expected in Surveys}

   Under the scenario that old white dwarfs make up the bulk of the Galactic dark
matter, they
will be plentiful but difficult to
detect because of their intrinsic faintness. The local number of such objects
can be determined simply from the mass density required to produce a flat Galactic rotation
curve  
($0.0079 M_{\odot}$/pc$^3$; Alcock {\it et al.} 1997a; Chabrier and M$\acute{e}$ra 1997; Gould, Flynn and Bahcall 1996) and the mean white dwarf mass $<M_{WD}>$. With the assumption that old white dwarfs are 100\%
of the Galactic dark matter, the 
local number density of these objects will then be 

\begin{displaymath} {Local~Number~WDs~/pc{^3}} = {\frac{0.0079} {<M_{WD}>}} .
\end{displaymath}

Richer {\it et al.} 1999 have constructed synthetic halo white dwarf luminosity functions for ages of 10, 14 and 16 Gyr
and limiting magnitude $V_{lim} = 28$ under the assumption that hydrogen-rich
white dwarfs make up 100\% of the Galactic dark matter. That is, the
functions were normalized by
the known local dark matter mass density.   
The following Table indicates some results from this work. 

\vspace{1cm} 
\centerline{\bf Table 1 - Predicted Number of White Dwarfs}

\begin{table}[h]
\hspace{1.5cm} 
\begin{tabular}{|l|c|c|c|}
\hline
\multicolumn{4}{|c|}{Number of White Dwarfs} \\
\hline
                &HDF: $V_{lim}=28$ (C)  &HDF: $V_{lim}=28$ (S) &Field: $V_{lim}=24$ (C)\\
\hline
10 Gyr         &   15   & 125   &  68    \\
14 Gyr         &  3   & 89  &  14\\
16 Gyr          &   2   &  92  &  10    \\
\hline
\end{tabular}
\end{table}

In this Table the second and third columns are the expected number of old white dwarfs
in the Hubble Deep Field (area 4.4 square arc minutes) to a limiting magnitude of $V_{lim} = 28$. These numbers were
calculated under 2 assumptions for the IMF of the white dwarf precursors; a
Chabrier-type IMF (C) and a Salpeter IMF (S). The last column is the
number of halo white dwarfs per square degree expected to a limiting
magnitude of $V_{lim} = 24$ using a Chabrier IMF. This will be approximately
the 100\% completion limit 
of deep ground-based surveys which are already in progress at CFHT, ESO and NOAO.

Some conclusions can already be drawn from Table 1. In particular, we can
eliminate the hypothesis that old white dwarfs with hydrogen-rich atmospheres, formed with a Salpeter
IMF make up the bulk of the Galactic dark matter. From the third column we
see that this would have yielded about 100 such objects in the HDF to 
$V_{lim} = 28$
 no matter what the
age of the halo. 
Ibata {\it et al.} 1999 have shown that there are at most 12 such candidates
in the HDF. The
suggestion that the dark matter consists of white dwarfs formed with a Chabrier IMF
cannot as yet be excluded by the observations.

For gound-based surveys the numbers are interestingly high - 68 are
expected per square degree to a limiting magnitude of $V_{lim} = 24$ for a 10
Gyr halo and about a dozen for a halo with an age of 14 -- 16 Gyr. It will thus prove to be quite straightforward to test the
white dwarf hypothesis with ground-based data and eventually even
use the statistics
to provide an accurate age for the halo. Discovery of such objects
from the ground will be critical as they will be bright enough that the current generation
of 8m class telescopes will be able to obtain spectra of them to confirm that
they are indeed old white dwarfs.

\acknowledgements
The author would like to thank the director and staff of the Osservatorio
Astronomico di Roma for providing support and a stimulating research
environment during his residency in Rome. The research of the author is supported, in
part, by the Natural Sciences and Engineering Research Council of Canada.


\end{document}